\documentclass[10pt]{article}

\usepackage{cite} 
\usepackage{amsmath}
\usepackage{amsfonts}
\usepackage{stmaryrd}
\usepackage{graphics} 
\usepackage{graphicx}
\usepackage{comment}
\usepackage{float}
\graphicspath{{Figures/}}
\usepackage{color}
\usepackage{authblk} 

\begin{document}
\title{Coupled Elastic-Acoustic Modelling for Quantitative Photoacoustic Tomography\\}
\author[1,2,3]{Hwan Goh}
\author[3]{Timo L\"{a}hivaara}
\author[4,6]{Tanja Tarvainen}
\author[4]{Aki Pulkkinen}
\author[1]{Owen Dillon}
\author[5]{Ruanui Nicholson}
\author[1,2,4]{Jari Kaipio}
\affil[1]{Department of Mathematics, University of Auckland, Private Bag 92019, Auckland Mail Centre, Auckland 1142, New Zealand.}
\affil[2]{Dodd-Walls Centre for Photonic and Quantum Technologies, University of Otago, P.O. Box 56, Dunedin 9056, New Zealand.}
\affil[3]{Oden Institute for Computational Engineering and Sciences, The University of Texas at Austin, Austin, TX 78712, USA.}
\affil[4]{Department of Applied Physics, University of Eastern Finland, P.O. Box 1627, 70211 Kuopio, Finland.}
\affil[5]{Department of Engineering Science, University of Auckland, Private Bag 92019, Auckland Mail Centre, Auckland 1142, New Zealand}
\affil[6]{Department of Computer Science, University College London, Gower Street, London WCIE 6BT, United Kingdom.}

\maketitle

\begin{abstract}
Quantitative photoacoustic tomography (qPAT) is an imaging technique aimed at estimating chromophore concentrations inside tissues from photoacoustic images, which are formed by combining optical information and ultrasonic propagation. The application of qPAT as a transcranial imaging modality is complicated by shear waves that can be produced when ultrasound waves travel from soft tissue to bone. Because of this, the estimation of chromophores distributions near the skull can be problematic. In this paper, we take steps towards compensating for aberrations of the recorded photoacoustic signals caused by elastic wave propagation. With photoacoustic data simulated in a coupled elastic-acoustic domain, we conduct inversions in a purely acoustic domain. Estimation of the posterior density of the initial pressure is achieved by inversion under the Bayesian framework. We utilize the Bayesian approximation error approach to compensate for the modelling errors arising from approximating a coupled elastic-acoustic domain with a purely fluid domain. The resulting reconstructions and corresponding uncertainty estimates are then used to evaluate the posterior density of the optical absorption parameter. In the sense of the posterior uncertainty, the results show that the Bayesian approximation error approach yields a more feasible estimate for the posterior model of the initial pressure which, in turn, yields a more feasible estimate for the posterior model of the absorption coefficient. 
\end{abstract}

\section{Introduction}
Photoacoustic tomography (PAT) is an imaging modality which combines the benefits of both optical contrast and spatial accuracy of ultrasound propagation \cite{xu2006photoacoustic,wang2008prospects,li2009photoacoustic,wang2009photoacoustic, beard2011biomedical, xia2014photoacoustic}. In PAT, a short pulse of near-infrared light, typically of nanoseconds duration, is used to illuminate the region of soft tissue of interest. For biological tissue, the incident energy from the light pulse is absorbed by light absorbing molecules within the tissue known as chromophores and converted into heat. The resulting rise in temperature induces rapid thermoelastic expansion of the chromophores, generating a localised increase in pressure. Due to the elastic nature of tissue, this local increase in pressure propagates through the tissue as an acoustic wave to be measured by ultrasound sensors on the surface of the tissue. This process is known as the photoacoustic effect. Soft tissue is usually highly optically scattering and so imaging to high resolution using purely optical means is difficult. In contrast, the scattering of acoustic waves of even up to tens of megahertz is considerably lower \cite{cox2007k}. Since the photoacoustic ultrasound waves carry this optical information directly to the surface with little scattering, accurate spatial information is retained.\par 
As such, PAT can be used to provide images of soft biological tissues with high spatial resolution. It has been successfully applied to the visualisation of different structures in biological tissues such as human blood vessels, microvasculature of tumours and cerebral cortex in small animals \cite{beard2011biomedical,xia2014small,razansky2012multispectral}. Specifically, the haemoglobin molecule is the primary absorber of light in blood and naturally facilitates photoacoustic imaging studies. Determining the properties of blood is of great interest because multiwavelength measurements of the absorption of blood provide functional information about tissue through oxygen saturation \cite{cox2012quantitative}.\par
There are two forms of haemoglobin molecules in blood: oxygenated and deoxygenated. Each possesses different absorption characteristics for light which results in optical contrast. With these properties, photoacoustic imaging can be applied to detecting brain hypoxia-ischemia cerebral injury \cite{cox2012quantitative}. However, the information provided by PAT is a qualitative image and it does not include quantitative information about chromophore concentration.\par
Whilst the inverse problem of PAT is to estimate the initial pressure distribution using time-resolved measurements of the propagating acoustic pressure waves recorded over the tissue surface, quantitative PAT (qPAT) is a technique in which the distribution of the optical parameters are estimated \cite{cox2012quantitative}. That is, the two inverse problems in qPAT are: reconstruction of the initial acoustic pressure distribution from measured acoustic waves and the following reconstruction of the distribution of the optical parameter from the absorbed optical energy density. As an alternative to the conventional two stage approach, estimation of the optical parameters directly from the photoacoustic measurements has been considered recently \cite{pulkkinen2016direct,haltmeier2015single,gao2015limited,ding2015one}. \par
The acoustic inverse problem of qPAT has been widely studied and there is a large number of reconstruction techniques available, see e.g. \cite{xu2006photoacoustic}, \cite{wang2009photoacoustic} and \cite{kuchment2008mathematics} and the references therein. The technical contribution in this paper is primarily motivated by using PAT as a transcranial imaging modality. {\em In vivo} transcranial PAT studies have been conducted in small animals to visualize brain structure, brain lesions and neurofunctional activities such as cerebral hemodynamic responses to hyperoxia and hypoxia as well as cerebral cortical responses to various forms of stimulation \cite{wang2003noninvasive,yang2007functional,li2010real,xu2011vivo}. 
However, shear waves can be produced when ultrasound waves travel from soft tissue to bone. In particular, the skull bone is a complex medium
that exhibits heterogeneities in the speed of sound, density, and attenuation which can distort an ultrasonic field. 
The skulls of small animals are relatively thin ($\sim$1 mm), and thus the photoacoustic waves may not be significantly aberrated by the bone.
The thickness of the human skull, however, will not only cause attenuation of illumination intensity and acoustic waves, but it also diffracts acoustic waves due to velocity mismatch and longitudinal-to-shear mode conversion \cite{li2009photoacoustic}. To this end, {\em ex vivo} studies involving primate heads have been conducted \cite{xu2006rhesus,jin2008effects,xu2011vivo,nie2011photoacoustic,huang2012aberration}. 
In such applications, the effects of the skull on the recorded photoacoustic wavefields are not negligible.
Therefore, to develop PAT as an effective imaging modality for transcanial imaging also in large primates, it is necessary to develop image reconstruction methodologies which accurately compensate for skull-induced aberrations of the recorded photoacoustic signals.\par
The acoustic properties of the skull have been modelled for transcranial ultrasound imaging applications \cite{fry1978acoustical,yousefi2009transcranial,hayner2001numerical,baikov2003physical}. Work has been done towards the goal of photoacoustic applications where the acoustic wave equation is utilized \cite{huang2012aberration,schoonover2011compensation}. Here, a fluid medium was assumed which yielded a simplified wave propagation model in which longitudinal-to-shear mode conversion was neglected. The deleterious effects of making such an approximation was studied in \cite{schoonover2012numerical}. Recently in \cite{mitsuhashi2017forward}, the skull was treated as an elastic solid through development of a numerical framework for image reconstruction in transcranial PAT based on an elastic wave equation.  The results showed significant improvement in the accuracy of reconstructed photoacoustic images when compared to reconstructions obtained ignoring the shear waves and using a back projection algorithm. \par
In this paper, we take steps towards developing image reconstruction methodologies to compensate for aberrations of the recorded photoacoustic signals caused by elastic wave propagation. We model illumination of the imaged target using a model for light transport and use the simulated photon fluence and absorbed optical energy density distributions to form the initial condition for the photoacoustic wave propagation which, in turn, is modelled in a coupled elastic-acoustic medium. With the simulated photoacoustic data, we conduct the acoustic inversion in a purely fluid domain. Estimation of the posterior density of the initial pressure is achieved by inversion under the Bayesian framework \cite{KaipioText,tarantola2005inverse,calvetti2007introduction}. We investigate the statistics of the approximation errors arising from approximating a coupled elastic-fluid media by a purely fluid domain. Furthermore, we utilize the Bayesian approximation error approach \cite{KaipioText} to marginalise over the elastic layer and improve the estimations. The resulting reconstructions and corresponding uncertainty estimates are then used to model the posterior density of the optical inverse problem which we solve using the Bayesian framework.. 

\section{Forward Problems}
The forward problem of qPAT aims to solve the photoacoustic time-series when optical and acoustic parameters and input light are given. The acoustic propagation itself occurs on a microsecond time scale which is approximately five orders of magnitude slower than the optical propagation. This large difference in time scales allows the optical and acoustic parts to be decoupled and treated separately \cite{li2009photoacoustic}.

\subsection{Optical forward problem}
In this work, propagation of light is modelled using the diffusion approximation (DA) to the radiative transfer equation \cite{arridge1999optical}. The DA with a Robin-type boundary condition is of the form
\begin{align}
&-\nabla \cdot \kappa({\bf x}) \nabla \Phi({\bf x}) + \mu_{\mathrm{a}}({\bf x}) \Phi({\bf x}) = 0,\:  {\bf x} \in \Omega \\
&\Phi({\bf x}) + \frac{1}{2\gamma_n}\kappa({\bf x})\frac{\partial \Phi({\bf x})}{\partial {\bf n}} = 
\begin{cases} 
\frac{I_{\mathrm{s}}}{\gamma_n}, & {\bf x}\in \epsilon_i,\\
0, & {\bf x} \in \partial \Omega \setminus \epsilon_i.
\end{cases}
\end{align}
In this form, we have domain $\Omega \subset \mathbb{R}^n$ with dimension $n = 2,3$, $\Phi$ is the photon fluence at ${\bf x} \in \Omega$ and $\mu_{\mathrm{a}}$ is the absorption coefficient. The term $\kappa = (n(\mu_{\mathrm{a}} + \mu_{\mathrm{s}}'))^{-1}$ is the diffusion coefficient with $\mu_{\mathrm{s}}' = (1-g)\mu_{\mathrm{s}}$ the reduced scattering coefficient, where $\mu_{\mathrm{s}}$ is the scattering coefficient and $g$ is the mean of the cosine of the scattering angle \cite{arridge1999optical,tarvainen2005coupled}. Further, $I_{\mathrm{s}}$ is a diffuse boundary current at the source position $\epsilon_j \in \partial \Omega$, $\widehat{n}$ is an outward unit normal and $\gamma_n$ a dimension-dependent constant which takes the values $\gamma_2 = \frac{1}{\pi}$ and $\gamma_3 = \frac{1}{4}$. In this work, a finite element model (FEM) is used to obtain the numerical solution $\Phi$ of the DA \cite{tarvainen2012reconstructing,tarvainen2005coupled}. The absorbed (light) energy density $h$ can then be solved from the fluence as
\begin{align}
h({\bf x}) = \mu_{\mathrm{a}}({\bf x})\Phi(\mu_{\mathrm{a}}({\bf x}),{\bf x}).
\end{align}
We connect the absorbed energy density to elastic wave propagation through use of the Duhamel-Neumann constitutive equation for the linear isotropic thermoelastic body \cite{Atanackovic}. In terms of the strain components, this is
\begin{align}
\varepsilon_{ij} = \frac{-\lambda \delta_{ij}}{2\mu(3\lambda + 2\mu)}\mathrm{tr}(\boldsymbol{\sigma}) + \frac{1}{2\mu}\sigma_{ij} + \alpha(T-T_0)\delta_{ij}
\end{align}
with $\boldsymbol{\varepsilon}$ is the strain tensor, $\boldsymbol{\sigma}$ is the stress tensor, $\lambda$ and $\mu$ are the first and second Lam\'{e} parameters, $\alpha$ is the linear thermal expansion coefficient, $\delta_{ij}$ is the Kronecker delta function, $T$ is the temperature and $T_0$ is the initial temperature. We assume free thermal expansion of the body by setting $\sigma_{ij} = 0$ and $(T - T_0) \neq 0$. This yields
\begin{align}
\varepsilon_{ij} = \alpha(T-T_0)\delta_{ij}.
\end{align}
In the context of qPAT, the temperature generated is $T - T_0 = \frac{h}{\rho C_v}$ where $\rho$ is the mass density of the tissue and $C_v$ is the specific heat capacity \cite{cox2009challenges}. This yields the initial condition for photoacoustic wave propagation in two dimensional elastic media:
\begin{align}
\boldsymbol{\varepsilon} ({\bf x},0) =
\begin{bmatrix}
\alpha({\bf x})\frac{h({\bf x})}{\rho({\bf x})C_v({\bf x})} & 0\\
0 & \alpha({\bf x})\frac{h({\bf x})}{\rho({\bf x})C_v({\bf x})}
\end{bmatrix}. \label{EqInitialStrain}
\end{align}
In our work, we assume $\alpha = 1$ and $C_v = 1$.

\subsection{Acoustic forward problem}
In this paper, we will use the term `acoustic forward problem' to refer to photoacoustic wave propagation in acoustic, elastic and coupled acoustic-elastic media. One of the governing equations for photoacoustic wave propagation is the strain-velocity formulation for the two dimensional isotropic elastic wave equation in the form of a first-order hyperbolic system
\begin{align}
\frac{\partial \boldsymbol{\varepsilon}({\bf x},t)}{\partial t} &= \frac{1}{2}(\nabla {\bf v}({\bf x},t) + \nabla {\bf v}({\bf x},t)^{\mathrm{T}}) \\
\rho({\bf x}) \frac{\partial {\bf v}({\bf x},t)}{\partial t} &= \nabla \cdot (\lambda({\bf x}) \mathrm{tr}(\boldsymbol{\varepsilon}({\bf x},t)){\bf I} + 2\mu({\bf x}) \boldsymbol{\varepsilon}({\bf x},t))
\end{align}
where, through the definition of the Lagrange-Green strain tensor, ${\bf v}$ is the velocity \cite{Atanackovic}. The corresponding conservative form is as follows
\begin{align}
{\bf Q}({\bf x}) \frac{\partial {\bf q}({\bf x},t)}{\partial t} + \nabla \cdot {\bf F}({\bf q}({\bf x},t)) = {\bf 0} \label{AdjStEWEDGMEqConsFormTens}
\end{align} 
where
\begin{align}
{\bf q}({\bf x},t) =
\begin{bmatrix}
\boldsymbol{\varepsilon}({\bf x},t)\\
{\bf v}({\bf x},t)
\end{bmatrix},\:
{\bf Q}({\bf x}) =
\begin{bmatrix}
{\bf I} & {\bf 0}\\
{\bf 0} & \rho({\bf x}){\bf I}
\end{bmatrix}
\end{align}
and ${\bf I}$ denoting the identity tensor, ${\bf 0}$ denoting the zero tensor and the flux operator ${\bf F}$ on the strain-velocity unknowns ${\bf q}$ is defined by
\begin{align}
({\bf F} ({\bf q}))_i = 
\begin{bmatrix}
-\frac{1}{2}({\bf v} \otimes {\bf e}_i + {\bf e}_i \otimes {\bf v})\\
-({\bf C}\boldsymbol{\varepsilon}){\bf e}_i
\end{bmatrix}
\end{align}
for $i \in \{1,2\}$ where $\otimes$ is the tensor product dropping the spatial and temporal dependence $({\bf x},t)$ for brevity. The divergence of the flux operator is then
\begin{align}
\nabla \cdot {\bf F} ({\bf q}) = -
\begin{bmatrix}
\frac{1}{2}(\nabla {\bf v} + \nabla {\bf v}^{\mathrm{T}})\\
\nabla \cdot (\lambda \mathrm{tr}(\boldsymbol{\varepsilon}){\bf I} + 2\mu \boldsymbol{\varepsilon})
\end{bmatrix}.
\end{align} 
We solve for the velocities using the discontinuous Galerkin method \cite{Wilcox,Hesthaven,lahivaara2018deep,schoeder2018optoacoustic}. The representation of the isotropic elastic wave equation as a first-order hyperbolic system allows the numerical flux to be constructed as a solution of a naturally induced Riemann problem over the interface of an element. This, in turn, allows the modelling of wave propagation in coupled elastic-acoustic media.\par
In order to formulate the discontinuous Galerkin scheme for the conservative form (\ref{AdjStEWEDGMEqConsFormTens}), we will use an inner product for the direct sum space ${\bf V} = V_{\mathrm{sym}}^{2 \times 2} \oplus V^2$ with the property that, for a test function $\boldsymbol{\phi}$ such that
\begin{align}
\boldsymbol{\phi} = 
\begin{bmatrix}
{\bf H}\\
{\bf w}
\end{bmatrix},
\end{align}
the inner product of $\boldsymbol{\phi}$ and ${\bf q}$ is
\begin{align}
{\bf q} \cdot \boldsymbol{\phi} = \boldsymbol{\varepsilon}:{\bf C}{\bf H} + {\bf v}\cdot{\bf w}. \label{AdjStEWEDGMEqInnerProd}
\end{align}
The use of the fourth-order elasticity tensor ${\bf C}$ in (\ref{AdjStEWEDGMEqInnerProd}) eliminates redundant variables when acoustic waves are treated as a special case of elastic waves in a unified elastic-acoustic formulation \cite{Wilcox}. Without assuming a particular spatial discretization of the domain $\Omega$, the discontinuous Galerkin formulation over an element $\Omega^k \subset \Omega$ is then obtained by using integration by parts twice on the inner product of the conservative form (\ref{AdjStEWEDGMEqConsFormTens}) and a test function $\boldsymbol{\phi}$.
This yields the localised variational form
\begin{align}
&\int_{\Omega^k} {\bf Q}\frac{\partial {\bf q}}{\partial t} \cdot \boldsymbol{\phi} \, \mathrm{d}{\bf x} + \int_{\Omega^k} (\nabla \cdot {\bf F}({\bf q}))\cdot \boldsymbol{\phi}\, \mathrm{d}{\bf x} \nonumber\\
&+ \int_{\partial \Omega^k}({\bf n} \cdot (({\bf F}({\bf q})^{\star} - {\bf F}^-({\bf q}^-)))\cdot \boldsymbol{\phi}\, \mathrm{d}{\bf x} = 0 \label{EqNumFluxTerm}
\end{align}
where the superscript `-' denotes the interior of the element $\Omega^k$ and the superscript `$\star$' denotes the numerical flux across the element interfaces. Expanding the inner product in the domain integrals gives
\begin{align}
&\int_{\Omega^k} \frac{\partial \boldsymbol{\varepsilon}}{\partial t}:{\bf C}{\bf H}\, \mathrm{d}{\bf x} + \int_{\Omega^k}\rho \frac{\partial {\bf v}}{\partial t} \cdot {\bf w}\, \mathrm{d}{\bf x}\nonumber\\
&- \int_{\Omega^k}\frac{1}{2}(\nabla {\bf v} + \nabla{\bf v}^{\mathrm{T}}):{\bf C}{\bf H}\, \mathrm{d}{\bf x}\nonumber\\
&- \int_{\Omega^k} (\nabla \cdot ({\bf C}\boldsymbol{\varepsilon}))\cdot {\bf w}\, \mathrm{d}{\bf x}\nonumber\\
&+\int_{\partial \Omega^k} {\bf n} \cdot \left(\left({\bf F}
\begin{bmatrix}
\boldsymbol{\varepsilon}\\
{\bf v}
\end{bmatrix}\right)^{\star} - {\bf F}^-
\begin{bmatrix}
\boldsymbol{\varepsilon}^-\\
{\bf v}^-
\end{bmatrix} \right) \cdot
\begin{bmatrix}
{\bf C}{\bf H}^-\\
{\bf w}^-
\end{bmatrix}
\, \mathrm{d}{\bf x}\nonumber\\
& = 0. 
\end{align}\label{AdjStEWEDGMEqConsFormTensFormExp}
At the interface between two elements, we have the following continuity conditions on the traction and velocities:
\begin{align}
\boldsymbol{\sigma}^-{\bf n} = \boldsymbol{\sigma}^+{\bf n}, &\textrm{ on } \partial\Omega^k,\\
{\bf v}^- = {\bf v}^+, &\textrm{ on } \partial\Omega^k, \label{EqContEls}\\
{\bf n}\cdot{\bf v}^- = {\bf n}\cdot{\bf v}^+, &\textrm{ on } \partial\Omega^k \label{EqContElsAcs}
\end{align}
where (\ref{EqContEls}) holds at the interface between two elements in elastic media and (\ref{EqContElsAcs}) holds at the interface between two elements in elastic-acoustic media or acoustic media. 
The superscripts $+$ and $-$, respectively, denote the outward and inward limits in the direction of the normal vector ${\bf n}$ of the element $\Omega^k$. 
We also invoke stress-free boundary conditions for the imaged domain:
\begin{align}
\boldsymbol{\sigma}^+ = -\boldsymbol{\sigma}^-, &\textrm{ on } \partial\Omega.
\end{align}
Solving for the Riemann problem induced over the element interfaces yields the numerical flux term (\ref{EqNumFluxTerm}) as
\begin{align}
&{\bf n} \cdot (({\bf F}({\bf q})^{\star} - {\bf F}^-({\bf q}^-))\nonumber\\
&= k_0 ({\bf n}\cdot\llbracket \sigma \rrbracket + \rho^+c_p^+\llbracket {\bf v} \rrbracket)
\begin{bmatrix}
{\bf n} \otimes {\bf n}\\
c_p^- {\bf n}
\end{bmatrix}\nonumber\\
&- k_1
\begin{bmatrix}
\mathrm{sym({\bf n}\otimes ({\bf n}\times({\bf n}\times \llbracket \boldsymbol{\sigma} \rrbracket)))}\\
c_s^- {\bf n}\times({\bf n}\times \llbracket \boldsymbol{\sigma} \rrbracket)
\end{bmatrix}\nonumber\\
& - k_1\rho^+c_s^+
\begin{bmatrix}
\mathrm{sym({\bf n}\otimes ({\bf n}\times({\bf n}\times [{\bf v}])))}\\
c_s^- {\bf n}\times({\bf n}\times [{\bf v}])
\end{bmatrix} \label{EqFlux3}
\end{align}
where $\mathrm{sym}$ is the symmetry operator, $c_p$ and $c_s$ are the pressure and shear wave speeds with
\begin{align}
c_p = \sqrt{\frac{\lambda + 2\mu}{\rho}}, \: c_s = \sqrt{\frac{\mu}{\rho}}
\end{align}
and 
\begin{align}
\llbracket \sigma \rrbracket  &= \sigma^-{\bf n}^- + \sigma^{+}{\bf n}^{+}\\
\llbracket {\bf v} \rrbracket  &= {\bf n}^-\cdot{\bf v}^-  + {\bf n}^{+}\cdot{\bf v}^{+}\\
\left[{\bf v}\right] &= {\bf v}^- - {\bf v}^{+} \label{AdjStEWEWDGMEqStateDiffFluxDef}
\end{align}
where the terms $k_0$ and $k_1$ are chosen as
\begin{align}
k_0 &= \frac{1}{\rho^-c_p^- +\rho^+c_p^+},\\
k_1 &= 
\begin{cases}
\frac{1}{\rho^-c_s^- +\rho^+c_s^+} &\textrm{ if } \mu^- \neq 0,\\
0 &\textrm{ if } \mu^- = 0.
\end{cases}
\end{align}
Specifically, the latter term is responsible for yielding a unified expression of the upwind numerical flux for domains consisting of all types of acoustic and elastic interfaces \cite{Wilcox}.

\section{Inverse Problems}
The inverse problem in qPAT is to solve the distribution of optical parameters in the medium using the time series data of the velocities. We utilize the Bayesian approach to ill-posed inverse problems for both the acoustic and optical inverse problems.

\subsection{Bayesian approach} \label{SecBayesianApproach}
In the Bayesian approach, the inverse problem is viewed in the framework of statistical inference \cite{KaipioText,tarantola2005inverse}. All parameters are treated as random variables which depend on each other through a model and information about these parameters expressed by probability distributions. Thus, the Bayesian approach yields not only estimates of the parameters of interest but the estimates of their uncertainties as well. We briefly detail the Bayesian approach.\par
Suppose we have an inverse problem which tasks us with determining unknown parameter distribution $x \in \mathbb{R}^N$ given noisy measurements $y_{\mathrm{d}} \in \mathbb{R}^M$. Now consider an observational model with additive errors
\begin{align}
y_{\mathrm{d}} = f(x) + e, \label{EqObsModel}
\end{align}
where $f: \mathbb{R}^N \to \mathbb{R}^M$ is the forward model and $e \in \mathbb{R}^M$ is the random variable denoting the additive error or noise. The observation model (\ref{EqObsModel}) links the parameters of interest to the measurements. Within the Bayesian approach, the observation model is interpreted statistically by defining probability distributions for the unknown parameter $x$ and noise $e$. The probability distribution $\pi_x$ of $x$ is called the prior distribution and it is the model for the marginal distribution of the unknown $x$, while the noise statistics characterises the measurement setup and modelling errors. We denote the prior and the noise probability density functions as $\pi_x$ and $\pi_e$ respectively. Assuming that $x$ and $e$ are uncorrelated in the additive noise model (\ref{EqObsModel}), the Bayesian approach results in posterior distribution $\pi(x|y_{\mathrm{d}})$ for the unknown $x$ conditioned by the measurements $y_{\mathrm{d}}$, and is given by
\begin{align}
\pi(x|y_{\mathrm{d}} ) \propto \pi_e(y_{\mathrm{d}}  -f(x))\pi_x(x). \label{EqPostDistribution}
\end{align}
Even with the assumption of mutual independence that gave rise to the likelihood model, it may still be the case that no closed form exists for this equation. In such cases, statistical methods such as Markov Chain Monte Carlo methods could be used to numerically approximate these densities \cite{KaipioText}. However, these methods may be computationally too expensive in large dimensional tomographic inverse problems. Therefore, point estimates such as the {\em maximum a posteriori} (MAP) estimate are often computed. In this work, the distributions $\pi_x$ and $\pi_e$ are modelled as Gaussian distributions and their parameters are denoted with $x \sim \mathcal{N}(\eta_x,\Gamma_x)$ and $e \sim \mathcal{N}(\eta_e,\Gamma_e)$. With the Gaussian choice for distributions, the negative logarithm of the posterior distribution (\ref{EqPostDistribution}) becomes
\begin{align}
u(x) = \frac{1}{2}\|L_e(y_{\mathrm{d}}  - f(x) - \eta_e)\|^2 + \frac{1}{2}\|L_x(x - \eta_x)\|^2,
\end{align}
where $L_e$ and $L_x$ are such that $L_e^{\mathrm{T}}L_e = \Gamma_e^{-1}$ and $L_x^{\mathrm{T}}L_x = \Gamma_x^{-1}$. The point estimate of $x$ used in this paper is the MAP estimate
\begin{align}
x_{\mathrm{MAP}} = \arg \min_x u(x).
\end{align}
Further, the uncertainty estimates we use are the credibility intervals. We compute approximations for the credibility intervals that are based on a local Gaussian approximation of the posterior distribution at the MAP estimate. These approximations are computed as follows. The forward model $f$ is approximated by the first order Taylor series at $x_{\mathrm{MAP}}$
\begin{align}
f(x) \approx f(x_{\mathrm{MAP}}) + J_f(x_{\mathrm{MAP}})(x - x_{\mathrm{MAP}})
\end{align}
where $J_f(x_{\mathrm{MAP}})$ is the Jacobian matrix of $f$ evaluated at $x= x_{\mathrm{MAP}}$. Then, the Taylor series approximation is substituted into the observation model (\ref{EqObsModel}). By forming the mean and covariance matrix of the joint distribution of $(x,y_{\mathrm{d}} )$ and then using the Schur complements to obtain the conditional distribution of $x|y_{\mathrm{d}} $, we obtain the Gaussian approximation for the posterior distribution. That is, $x|y_{\mathrm{d}}  \sim \mathcal{N}(\eta, \Gamma)$ where the approximate posterior mean is $\eta = x_{\mathrm{MAP}}$ and the posterior covariance is
\begin{align}
\Gamma_{\mathrm{post}} = \left(J_f(x_{\mathrm{MAP}})^{\mathrm{T}}\Gamma_e^{-1}J_f(x_{\mathrm{MAP}}) + \Gamma_x^{-1}\right)^{-1}.
\end{align}
For true Gaussian distributions, 99.7\% of the probability mass of each element $x_i$ of $x$ would lie in the interval $\eta_i \pm 3\sigma_i$ where $\sigma_i = \sqrt{\Gamma_{\mathrm{post}}(i,i)}$. In this paper, we refer to these intervals as posterior error estimates. See \cite{pulkkinen2016direct} for a discussion on approximations of posterior error estimates in the context of qPAT. Next, we consider the Bayesian approximation error approach which can be used to account for modelling errors in the inversion process.
\subsection{Bayesian approximation error approach} \label{SecBAE}
The approximation error approach was introduced in \cite{KaipioText,KaipioAE}. Although this approach was initially motivated by an attempt to counteract the errors that arise from discretization, it can be used to account for modelling errors in general as well. For example, model reduction, domain truncation, unknown anisotropy structures and marginalization of uninteresting distributed parameters  in other optical and ultrasonic imaging modalities were treated in \cite{arridge2006approximation,kolehmainen2009approximation,heino2004modelling,heino2005compensation,kolehmainen2011marginalization,koponen2014bayesian,tarvainen2010approximation}. For the optical inverse problem of qPAT, this approach was used to account for the modelling errors. Furthermore, in \cite{tarvainen2013bayesian}, modelling of noise and erros due to the acoustic solver were considered when solving the optical inverse problem of qPAT. These errors may otherwise cause a significantly detrimental effect on the solution due to the ill-posed nature of inverse problems.\par
The consideration of inverse problems under the Bayesian paradigm allows us to account for computational model inaccuracies by representing these as an additional additive random variable. In this work, we use this approach for the approximate premarginalization of the uninteresting unknowns and uncertainties. Premarginalization refers to the act of marginalizing the uninteresting unknowns by eliminating them before inference. In this work, the uninteresting unknown that we premarginalize is the second Lam\'{e} parameter; the shear modulus responsible for elastic wave propagation through the skull. This allows for inversion in a purely fluid domain. We now briefly outline the details of this approach.\par
Consider again the finite-dimensional inverse problem with additive noise (\ref{EqObsModel}). Suppose that the noise and unknown are mutually independent Gaussian random variables. Suppose that the forward model $x \mapsto B(x)$ is used instead of $x \mapsto A(x)$. Then
\begin{align}
y_{\mathrm{d}} = B(x) + (A(x) - B(x)) + e. \label{EqApproxErrorFiniteAB}
\end{align} 
We may consider $A$ to represent the `accurate' model and $B$ a `less accurate' model; it is often the case that $B$ is considered to be less accurate because it is an approximation of $A$. Note that $A$ and $B$ may differ for properties other than discretization; the operator $B$ may lack something that $A$ possesses which renders it a less accurate approximation. Let us denote $\varepsilon(x) = A(x) - B(x)$. With this, (\ref{EqApproxErrorFiniteAB}) becomes
\begin{align}
y_{\mathrm{d}} = B(x) + \varepsilon(x) + e. \label{EqApproxErrorFiniteeps}
\end{align} 
where $\varepsilon(x)$ is the aforementioned additional additive random variable representing modelling errors and uncertainties. We note that (\ref{EqApproxErrorFiniteeps}) is exact. We approximate the statistics of $\epsilon$ using a normal distribution formed by sample means and sample covariances. To obtain this, we generate a set of samples $\{x^{(\ell)}: \ell = 1,\dots,L\}$. Then, we generate the sample approximation errors as
\begin{align}
\varepsilon^{(\ell)} = A(x^{(\ell)}) - B(x^{(\ell)})
\end{align}
which yields the sample means as
\begin{align}
\hat{\eta}_{\varepsilon} = \frac{1}{L} \sum_{\ell=1}^L \varepsilon^{(\ell)}
\end{align}
and sample covariances as
\begin{align}
\hat{\Gamma}_{\varepsilon} &= \frac{1}{L-1} \sum_{\ell=1}^L \varepsilon^{(\ell)}\varepsilon^{(\ell)\mathrm{T}} - \frac{L}{L-1}\hat{\eta}_{\varepsilon} \hat{\eta}_{\varepsilon}^{\mathrm{T}}.
\end{align} 
Then, with the assumption that $\varepsilon$ and $x$ are mutually independent, we obtain an approximation to the posterior distribution known as the enhanced error model \cite{KaipioText}. With the Gaussian approximation for distributions, the negative logarithm of the enhanced error model is
\begin{align}
u_{\mathrm{enh}}(x) = \frac{1}{2}\|L_v(y_{\mathrm{d}}  - B(x) - \eta_v)\|^2 + \frac{1}{2}\|L_x(x - \eta_x)\|^2
\end{align}
where $\eta_v = \eta_e + \hat{\eta}_{\varepsilon}$ and $L_v$ is such that $\Gamma_v = \Gamma_e + \hat{\Gamma}_{\varepsilon}$.\par
We now apply this discussion to both the acoustic and optical inverse problems. For the latter, we follow \cite{tarvainen2013bayesian} where the Bayesian approach was utilized. For the acoustic inverse problem, however, in \cite{tarvainen2013bayesian} a time reversal method utilizing the k-Wave MATLAB toolbox \cite{treeby2010k} was used in the deterministic framework. In \cite{tick2016image}, the Bayesian approach with the k-space time-domain method used as the numerical solution of linear wave equation implemented with the k-Wave MATLAB toolbox was considered. In this paper, we use the Bayesian approach with the discontinuous Galerkin method. Before we proceed to discussing the details, we will first briefly mention the prior model we used for both acoustic and optical inverse problems.
\subsection{Prior model} \label{SecPriorModel}
In this work, the prior model for the unknown initial pressure and unknown absorption coefficient were both informative smoothness priors \cite{KaipioText,arridge2006approximation,kolehmainen2009approximation}.
The informative smoothness prior is based on modifying a smoothing preprior into a proper Gaussian distribution. The smoothing preprior we use possesses a sparse covariance matrix that is motivated by a Gaussian Markov random field model \cite{KaipioText,Rue}. In this approach, quantitative information of the properties of the discretised parameter at marginalization points is included into the smoothing preprior. This information represents the spatial characteristic length scale for the estimated parameters which is roughly the prior estimate of the spatial size of the inhomogeneities in the target domain. In this work, the correlation length for both the initial pressure and absorption coefficient is set at $4$mm.

\subsection{Acoustic inverse problem}
Usually, the acoustic inverse problem is to reconstruct the initial acoustic pressure distribution from measured acoustic waves. However, in the setting of coupled elastic-acoustic wave propagation, the mass density is inhomogeneous. Therefore, in our work, the acoustic inverse problem instead is to reconstruct the absorbed energy density from the measured velocity of the photoacoustic waves. Thus, the medium density is treated as a component of the acoustic forward mapping from the absorbed energy density to the velocity data.\par
We denote our vectorized simulated time series velocity data as ${\bf v}_{\mathrm{d}} = [{\bf v}_{1,\mathrm{d}},{\bf v}_{2,\mathrm{d}}]^{\mathrm{T}} \in \mathbb{R}^{2n_sn_t}$ where $n_s$ denotes the number of sensors and $n_t$ the number of discretised time steps. As discussed in Section \ref{SecBayesianApproach}, with the Gaussian choice for distributions, the negative logarithm of the posterior distribution (\ref{EqPostDistribution}) becomes
\begin{align}
u(h) = \frac{1}{2}\|L_e({\bf v}_{\mathrm{d}} - f(h) - \eta_e)\|^2 + \frac{1}{2}\|L_h(h - \eta_h)\|^2,
\end{align}
where the absorbed energy density $h$ is as in (\ref{EqInitialStrain}). Thus, our parameter of interest for the acoustic inverse problem is a component of the initial condition of the acoustic forward problem. In component form, this is
\begin{align}
{\bf q}({\bf x},0) = 
\begin{bmatrix}
\varepsilon_{11}({\bf x},0)\\
\varepsilon_{22}({\bf x},0)\\
\varepsilon_{12}({\bf x},0)\\
v_1({\bf x},0)\\
v_2({\bf x},0)\\
\end{bmatrix}
=
\begin{bmatrix}
\frac{h({\bf x})}{\rho}\\
\frac{h({\bf x})}{\rho}\\
0\\
0\\
0\\
\end{bmatrix}
\end{align}
where we omit the variables $\alpha$ and $C_v$ as they are both set to $1$. To ensure our estimates of $h$ are positive, we implement a soft positivity constraint parameterizing $h$ using $z$ as
\begin{align}
h(z) = \frac{1}{\beta}\log(\exp(\beta z)+1)
\end{align}
where we set $\beta = 0.08$. To compute the MAP estimate of $h$, we utilize the Gauss-Newton algorithm \cite{Nocedal}. However, due to the non-linearity induced by the positivity constraint and the time dependency of the forward mapping, constructing the Jacobian matrix required for the Gauss-Newton algorithm through perturbations will be dependent on the dimension of the parameter space. Therefore, we instead utilize the adjoint-state method to form the Gauss-Newton approximation of the reduced Hessian. It can be shown that this requires only one forward solve and one adjoint solve; the latter being of similar computational complexity as a forward solve \cite{Gunzburger,Hinze,Troltzsch,FICHTNER200686,epanomeritakis2008newton,bui2013computational,petra2011model}. In doing so, we have a means of computing the search direction at each iteration which is independent of the dimension of the parameter space.\par
To do this for our case, we consider the Lagrangian 
\begin{align}
\mathcal{L}({\bf q},z,{\bf b},{\bf d}) &= \mathcal{J}({\bf q},z) + F({\bf q})({\bf b}) + G({\bf q},h(z))({\bf d})
\end{align}
where, in the infinite dimensional setting of function spaces, $\mathcal{J}({\bf q},z)$ can be written in the form
\begin{align}
&\frac{1}{2}\sum_{\chi=1}^{n_{\mathrm{s}}}\int_0^{t_f}\int_{\Omega}L_e\left({\bf v}_{\mathrm{d}} -{\bf v}\right)\cdot L_e\left({\bf v}_{\mathrm{d}} -{\bf v}\right) \delta({\bf x} - {\bf s}_{\chi}) \, \mathrm{d}{\bf x}\mathrm{d}t \nonumber\\
&+ \frac{1}{2}\int_{\Omega} L_z(\eta_{z} - z)^2 \, \mathrm{d}{\bf x},
\end{align} 
where ${\bf s}_{\chi} \in \Omega$ denotes the sensor on the boundary of the domain and $t_f$ denotes the final time. Also, $F({\bf q})({\bf b})$ can be written in the form
\begin{align}
&\int_0^{t_f} \sum_k\bigg(\int_{\Omega^k} {\bf Q}\frac{\partial {\bf q}}{\partial t} \cdot {\bf b} \, \mathrm{d}{\bf x} + \int_{\Omega^k} (\nabla \cdot {\bf F}({\bf q}))\cdot {\bf b}\, \mathrm{d}{\bf x} \nonumber\\
&+ \int_{\partial \Omega^k}({\bf n} \cdot (({\bf F}({\bf q})^{*} - {\bf F}^-({\bf q}^-)))\cdot {\bf b}\, \mathrm{d}{\bf x}\bigg)\, \mathrm{d}t 
\end{align}
and $G({\bf q},h(z))({\bf d})$ can be written in the form
\begin{align}
\int_{\Omega}
\left(
\begin{bmatrix}
\varepsilon_{11}({\bf x},0)\\
\varepsilon_{22}({\bf x},0)\\
\varepsilon_{12}({\bf x},0)\\
v_1({\bf x},0)\\
v_2({\bf x},0)
\end{bmatrix}
-
\begin{bmatrix}
\alpha({\bf x})\frac{h(z)({\bf x})}{\rho({\bf x})C_v({\bf x})}\\
\alpha({\bf x})\frac{h(z)({\bf x})}{\rho({\bf x})C_v({\bf x})}\\
0\\
0\\
0
\end{bmatrix}\right) \cdot {\bf d} \, \mathrm{d}{\bf x}.
\end{align}
Here, the variables ${\bf b}$ and ${\bf d}$ are the Lagrange multipliers which are also called the adjoint variables. Forming the Gauss-Newton approximation of the reduced Hessian necessitates the partial differentiation of the Lagrangian with respect to the state variable ${\bf q}$, the parameter of interest $h$ and the Lagrange multipliers. Differentiation with respect to the state variable ${\bf q}$ yields the adjoint equation
\begin{align}
F_{\bf q}'[{\bf q}](\tilde{\bf q})({\bf b}) + G_{\bf q}'[{\bf q},h(z)](\tilde{\bf q})({\bf d}) = -\mathcal{J}_{\bf q}'[{\bf q},h(z)](\tilde{\bf q})
\end{align}
for which we require the adjoint of the Fr\'{e}chet derivative of the numerical flux (\ref{EqFlux3}). Denoting the adjoint variable as ${\bf b} = [\boldsymbol{\gamma},{\bf w}]^{\mathrm{T}}$, this is as given in \cite{WilcoxAdjoint}:
\begin{align}
&{\bf n} \cdot (({\bf F}_{\bf q}'[\bf q]^*({\bf b}))^{\star} - {\bf F}_{\bf q}'[\bf q]^{*-}({\bf b}^-)) =\nonumber\\
&k_0 ({\bf n}\cdot\llbracket \boldsymbol{\gamma} \rrbracket - \rho^+c_p^+\llbracket {\bf w} \rrbracket)
\begin{bmatrix}
-{\bf n} \otimes {\bf n}\\
c_p^- {\bf n}
\end{bmatrix} \nonumber\\
&- k_1
\begin{bmatrix}
-\mathrm{sym({\bf n}\otimes ({\bf n}\times({\bf n}\times \llbracket \boldsymbol{\gamma} \rrbracket)))}\\
c_s^- {\bf n}\times({\bf n}\times \llbracket \boldsymbol{\gamma} \rrbracket)
\end{bmatrix}\nonumber\\
&- k_1\rho^+c_s^+
\begin{bmatrix}
\mathrm{sym({\bf n}\otimes ({\bf n}\times({\bf n}\times [{\bf w}])))}\\
-c_s^- {\bf n}\times({\bf n}\times [{\bf w}])
\end{bmatrix}. \label{EqAdjFlux3}
\end{align}
\subsection{Optical inverse problem} \label{SecOpticalInverseProblem}
The optical inverse problem is to reconstruct the distributions of the optical parameters from the absorbed optical energy density; the latter obtained from the reconstructed initial pressure estimated through the acoustic inverse problem. With the Gaussian choice for distributions, the negative logarithm of the posterior distribution becomes
\begin{align}
u(\mu_{\mathrm{a}}) &= \frac{1}{2}\|L_{e}(h_{\mathrm{MAP}}-\mu_{\mathrm{a}} \Phi(\mu_{\mathrm{a}}) - \eta_e)\|^2 \nonumber\\
&+ \frac{1}{2}\|L_{\mu_{\mathrm{a}}}(\mu_{\mathrm{a}}-\eta_{\mu_{\mathrm{a}}})\|^2 \label{mu_aPost}.
\end{align}
Since the solution of the acoustic inverse problem serves as the data for the optical inverse problem, then, to form the error model, we consider the statistics of the error of this data. We construct this error model now.\par
Since our data is a distribution, as obtained using in the acoustic inverse problem, we can express the forward model as
\begin{align}
y_{\mathrm{d}} - e = \mu_{\mathrm{a}} \Phi(\mu_{\mathrm{a}}),
\end{align}
where $y_{\mathrm{d}} - e \sim \mathcal{N}(h_{\mathrm{MAP}},\Gamma_{h,\mathrm{post}})$, with $y_{\mathrm{d}} = h_{\mathrm{MAP}}$ and $e \sim N(0, \Gamma_{h,\mathrm{post}})$. This can then be expressed in the traditional form of (\ref{EqObsModel}) simply as
\begin{align}
h_{\mathrm{MAP}} = \mu_{\mathrm{a}} \odot \Phi(\mu_{\mathrm{a}}) + e. \label{EqObservationModelOptInv2}
\end{align}
Thus, in reference to (\ref{mu_aPost}), we are motivated by (\ref{EqObservationModelOptInv2}) to set $\eta_e = 0$ and $L_{e}$ such that $L_{e}^{\mathrm{T}}L_{e} = \Gamma_{h,\mathrm{post}}^{-1}$. We call this the coupled error model (CEM). With this model, we obtain the MAP estimate of $\mu_{\mathrm{a}}$ using again the Gauss-Newton algorithm. However, due to the time independence of the optical forward mapping, there is no need to employ the adjoint-state method to construct the Gauss-Newton search direction. Instead, we directly construct the search direction through construction of the Jacobian of the finite element discretization of the mapping $\mu_{\mathrm{a}} \mapsto \mu_{\mathrm{a}} \odot \Phi(\mu_{\mathrm{a}})$.\par
In this paper, we simply set the scattering coefficient as a constant value $\mu_{\mathrm{s}} = 2\mathrm{mm}^{-1}$ in the domain; and so the absorption coefficient is the only reconstructed variable. Our choice of the constant value $\mu_{\mathrm{s}}$ is set to be equal to the true background value in the inversion process. It was shown in \cite{kolehmainen2011marginalization} that, in the case where a less accurate value is selected, the approximation error approach can be used to account for such inaccuracies to some extent. In \cite{pulkkinen2014approximate}, this was done in the context of qPAT.

\section{Results}
We tested the approach with numerical simulations. We obtain two posterior densities for the absorbed energy density $h$; one using the conventional noise model (CNM) and the other using the enhanced error model (EEM). Then, using each of those posterior densities, we obtain the posterior densities for the absorption coefficient $\mu_{\mathrm{a}}$ using the coupled error model yielding four posterior densities in total for the optical inverse problem.
\subsection{Computational domain and data simulation}
We considered a $\Omega = [-20\mathrm{mm},20\mathrm{mm}] \times [-20\mathrm{mm},20\mathrm{mm}]$ square domain. For the optical forward problem with the finite element discretization, we use a mesh consisting of $9300$ elements and $4785$ nodes to simulate the measurement data and, for inversion, we use a mesh consisting of $1706$ elements and $912$ nodes. For modelling the approximation error in the acoustic inverse problem, we use a mesh consisting of $8310$ elements and $4282$ nodes. The illumination sources are located along all four edges of the boundary.\par
\begin{figure}[H]
	\centering
	\includegraphics[width=4.3cm, height=4.3cm]{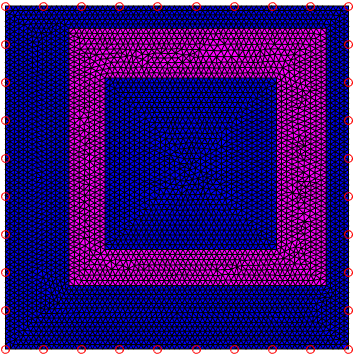}
	\includegraphics[width=4.3cm, height=4.3cm]{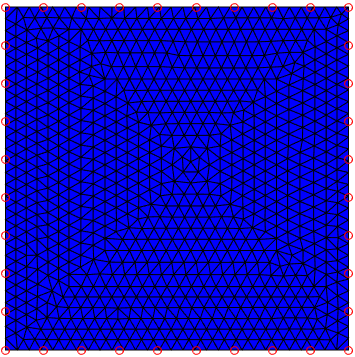}
	\caption{Top row: The two triangulated meshes used in the optical forward and inverse problem. Left: the data mesh. Right: the inversion mesh. The red circles represent the sensors. Fluid media is represented in blue and the elastic layer in magenta.} \label{FigureDataandInverseMesh}
\end{figure}

For the acoustic forward problem with discontinuous Galerkin discretization, we use a second order polynomial basis for both data simulation and inversion so that each element possesses 6 nodes. This corresponds to $55800$, $10236$ and $49860$ nodes for the data simulation mesh, inversion mesh  and the approximation error modelling mesh respectively. We use $36$ sensors located along the four edges of the domain. We simulate data from a fluid domain with an elastic layer. The triangulated data and inverse mesh with corresponding fluid and elastic medium are displayed in Figure \ref{FigureDataandInverseMesh}. 
The medium densities and wave speeds selected for the acoustic and elastic domain follow those used in \cite{mitsuhashi2017forward}. For the fluid medium, we set the medium density to be $\rho_f = 1000\frac{\mathrm{kg}}{\mathrm{m}^3}$ and the wave speed $c_f = 1500\frac{\mathrm{m}}{\mathrm{s}}$. For the elastic medium, we set the medium density to $\rho_e = 1850\frac{\mathrm{kg}}{\mathrm{m}^3}$ and the pressure and shear wave speeds to $c_p = 3000\frac{\mathrm{m}}{\mathrm{s}}$ and $c_s = 1500\frac{\mathrm{m}}{\mathrm{s}}$ respectively. We assume that both mediums are non-absorbing. The acoustic forward problem is temporally discretized into $743$ time steps, each $5.386 \mathrm{ns}$.  Forward computations were executed using MATLAB version R2014b on a Intel Core i5-4590 processor operating at $3.30 \mathrm{GHz}$. Computing $743$ time steps of the forward wave propagation on the data mesh takes approximately $90$ seconds.
\subsection{Measurement noise}
The noise afflicting the data is drawn from a Gaussian distribution with zero mean and standard deviation 
\begin{align}
\sigma_{v_\tau}(\chi) = \epsilon\left(\max_{t \in [0\mathrm{s},4\times 10^{-6}\mathrm{s}]}\left\{\max_{\chi \in \mathcal{S}}
\{|v_{\tau,\mathrm{d}}(\chi,t)|\}\right\}\right) \label{EqDataNoise}
\end{align}
with $v_{\tau,\mathrm{d}}(\chi,t)$ denoting the sensory data at sensor $\chi$ at time $t$ for $\tau \in \left\{1,2\right\}$ and $\mathcal{S}$ denoting the set of sensors. Further, we set $\epsilon = 0.05,0.01,0.001$ which corresponds to a noise levels of $5\%,1\%,0.1\%$ respectively.

\subsection{Acoustic inverse problem}
For the acoustic inverse problem, we consider two error models: the conventional noise model and the enhanced error model as detailed in Section \ref{SecBAE}. For the conventional noise model, we set $\eta_e = 0$ and $ \Gamma_e = \sigma_{e,\tau}^2I$ where $\sigma_{e,\tau}$ matches $\sigma_{v_\tau}$ in (\ref{EqDataNoise}). With this, it is clear that we are assuming the scenario that the statistics of the noise is accurately known.\par
\begin{figure}[H]
	\centering
	\includegraphics{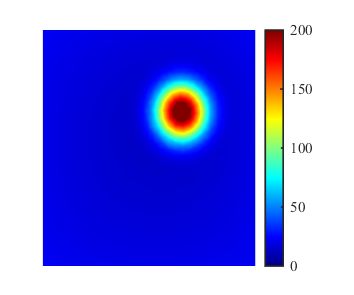}
	\caption{The true absorbed energy density distribution.}
	\label{FigureTrueAEDensity}
\end{figure}
\begin{figure}[H]
	\centering
	\begin{tabular}{c c c}
		\centering
		\includegraphics{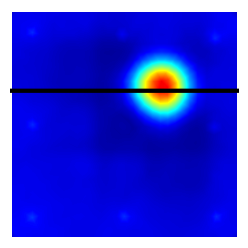} & 
		\includegraphics{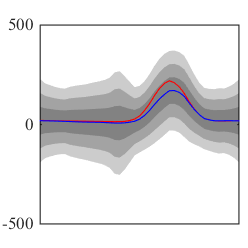}\\
		\includegraphics{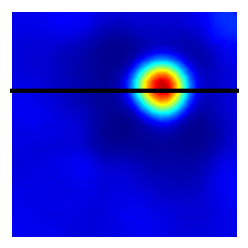} &
		\includegraphics{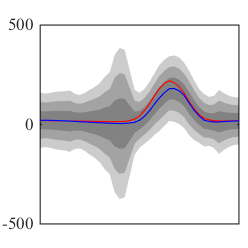}\\
		\includegraphics{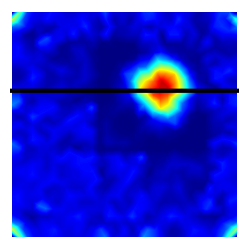}&
		\includegraphics{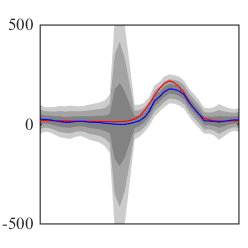}
	\end{tabular} 
	\caption{Conventional noise model posterior model for $h$. Left column: conventional noise model MAP estimates $h_{\mathrm{MAP}}^{\mathrm{CNM}}$. Right column: uncertainty estimates $\Gamma_{h,\mathrm{post}}^{\mathrm{CNM}}$ with 1, 2 and 3 standard deviations displayed by the shading. The red line represents the true values and the blue line represents the MAP estimate values. Rows: noise levels of $5\%,1\%,0.1\%$ respectively.}
	\label{FigureConvNoiseRecons}
\end{figure}
\begin{figure}[H]
	\centering
	\begin{tabular}{c c c}
		\centering
		\includegraphics{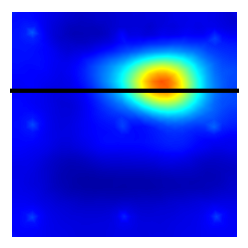} & 
		\includegraphics{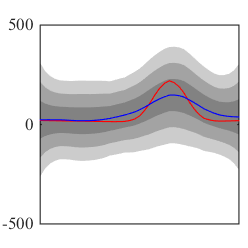}\\
		\includegraphics{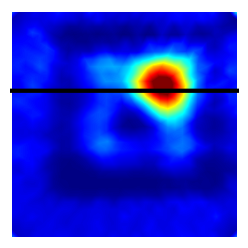} &
		\includegraphics{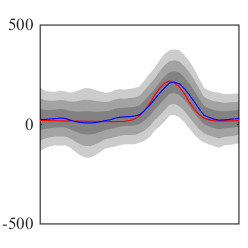}\\
		\includegraphics{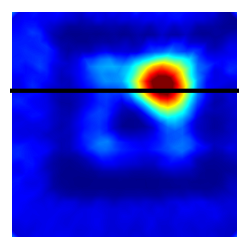}&
		\includegraphics{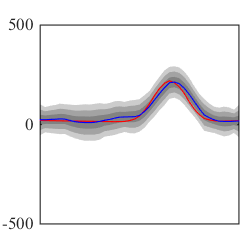}
	\end{tabular} 
	\caption{Enhanced error model posterior model for $h$. Left column: enhanced error model MAP estimates $h_{\mathrm{MAP}}^{\mathrm{EEM}}$. Right column: uncertainty estimates $\Gamma_{h,\mathrm{post}}^{\mathrm{EEM}}$ with 1, 2 and 3 standard deviations displayed by the shading. The red line represents the true values and the blue line represents the MAP estimate values. Rows: noise levels of $5\%,1\%,0.1\%$ respectively.}
	\label{FigureEnhNoiseRecons}
\end{figure}

We display the MAP estimates $h_{\mathrm{MAP}}^{\mathrm{CNM}}$, $h_{\mathrm{MAP}}^{\mathrm{EEM}}$ and uncertainty estimates $\Gamma_{h,\mathrm{post}}^{\mathrm{CNM}}$, $\Gamma_{h,\mathrm{post}}^{\mathrm{EEM}}$ for the two error models in Figures \ref{FigureConvNoiseRecons} and \ref{FigureEnhNoiseRecons} respectively. For each of the inversion cases, we also consider three noise levels: 5\%, 1\% and 0.1\%. Our metric for error is the relative error
\begin{align}
E_{h} = 100\%\cdot \frac{\|h_{\mathrm{true}} - h_{\mathrm{MAP}}\|}{\|h_{\mathrm{true}}\|} 
\end{align}
where $h_{\mathrm{MAP}}$ denotes the MAP estimate of $h$ and $h_{\mathrm{true}}$ denotes the true absorbed energy density shown in Figure \ref{FigureTrueAEDensity}. In Table \ref{TableAcousticInverseErrors}, we display the relative errors for the two error models we investigated at the three selected noise levels.\par
We begin the discussion of Figures \ref{FigureConvNoiseRecons} and \ref{FigureEnhNoiseRecons} by comparing between noise levels. For both figures, we can see that with a noise level of 5\%, both MAP estimates $h_{\mathrm{MAP}}^{\mathrm{CNM}}$ and $h_{\mathrm{MAP}}^{\mathrm{EEM}}$ have underestimated the true values of the absorbed energy density more compared to the MAP estimates with noise levels of 1\% and 0.1\%. They are smoother and the MAP estimates are closer to the prior mean $h(\mathbb{E}[Z_{\mathrm{pr}}]) \approx 70$.\par
Now we compare the MAP estimates $h_{\mathrm{MAP}}^{\mathrm{CNM}}$ and $h_{\mathrm{MAP}}^{\mathrm{EEM}}$ of the two error models. The conventional noise model MAP estimates $h_{\mathrm{MAP}}^{\mathrm{CNM}}$ in Figure \ref{FigureConvNoiseRecons} show some artefacts at the corners of the domain for the noise levels of 1\% and 0.1\% which are not present for the noise level of 5\%. This is due to the action of the smoothness prior which penalizes these artefacts in the inversion process when the data misfit is weighted lower by the noise model. Furthermore, we can see a dip in the background values resembling the shape of the interior of the elastic layer in the background values. For the enhanced error model results displayed in Figure \ref{FigureEnhNoiseRecons}, the background absorbed energy density values of $h_{\mathrm{MAP}}^{\mathrm{EEM}}$ are smoother and the corner artefacts are no longer present. We also note that at 5\% noise, the MAP estimates $h_{\mathrm{MAP}}^{\mathrm{EEM}}$ underestimate the values even more than when the conventional noise model was used. This may be because the enhanced error model covariance pushes our MAP estimates more towards the prior.\par
Now we discuss the uncertainty estimates $\Gamma_{h,\mathrm{post}}^{\mathrm{CNM}}$ and $\Gamma_{h,\mathrm{post}}^{\mathrm{EEM}}$ for both error models. In both Figures \ref{FigureConvNoiseRecons} and \ref{FigureEnhNoiseRecons}. The first thing we observe is that as the noise level decreases, the width of the error intervals decreases. This is to be expected as the data has more weight in the inversion process when the noise level of the noise model is set to be lower. Also, the true values sit comfortably within three standard deviation of the MAP estimate; thereby confirming that the estimates are feasible. For the conventional noise model MAP uncertainty estimates $\Gamma_{h,\mathrm{post}}^{\mathrm{CNM}}$, there is a noticeable spike in the uncertainty around the centre of the domain. This is especially pronounced for the noise levels of 1\% and 0.1\%. We speculate that the spike represents the uncertainty in the inversion process towards the centre of the domain within the elastic layer. In contrast, the enhanced error model error uncertainty estimates $\Gamma_{h,\mathrm{post}}^{\mathrm{EEM}}$ displayed in Figure \ref{FigureEnhNoiseRecons} do not exhibit this strange structure; perhaps because the elastic layer has been accounted for and therefore there is less uncertainty for the values within the elastic layer. Overall, we can see that using the enhanced error model yields a more accurate posterior model. This observation is also quantitatively supported by the relative errors displayed in Table \ref{TableAcousticInverseErrors}.
\begin{table}
	\renewcommand{\arraystretch}{1.3}
	\caption{Relative errors for MAP estimates from the acoustic inverse problem.}
	\label{TableAcousticInverseErrors}
	\centering
	\begin{tabular}{c | c | c | c}
	\centering
	& 5\% & 1\% & 0.1\% \\
	\hline
	Conventional: $h_{\mathrm{MAP}}^{\mathrm{CNM}}$\ & 33.1616 & 31.8243 & 33.6926\\
	Enhanced: $h_{\mathrm{MAP}}^{\mathrm{EEM}}$ & 30.5839 & 28.0243 & 24.0771 
\end{tabular}
\end{table}

\subsection{Optical inverse problem}
\begin{figure}[H]
	\centering
	\includegraphics{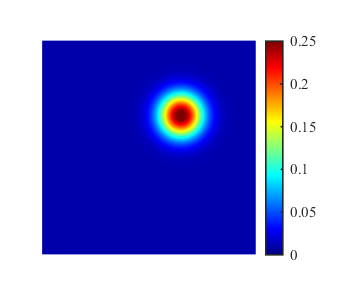}
	\caption{The true absorption coefficient distribution.}
	\label{FigureTruemu_a}
\end{figure}

For the optical inverse problem, we utilize the conventional noise model and the coupled error model as detailed in Section \ref{SecOpticalInverseProblem}. For both cases, the conventional noise model $h_{\mathrm{MAP}}^{\mathrm{CNM}}$ and enhanced error model $h_{\mathrm{MAP}}^{\mathrm{EEM}}$ MAP estimates of $h$ acts as data. For the case of the coupled error model, as discussed in Section \ref{SecOpticalInverseProblem}, the posterior covariances  $\Gamma_{h,\mathrm{post}}^{\mathrm{CNM}}$ and $\Gamma_{h,\mathrm{post}}^{\mathrm{EEM}}$ are also utilized as the error model covariance. For the conventional noise model, we set $\eta_e = 0$ and $\Gamma_e = \sigma_{e}^2 I$ where $I$ is the $912 \times 912$ identity matrix and $\sigma_{e} = \epsilon\max(h_{\mathrm{MAP}})$ with $\epsilon = 0.05, 0.01, 0.001$ corresponding to the anticipated noise level of the noise afflicting the photoacoustic data. For the coupled error model, we use the posterior covariance of $h$, $\Gamma_{h,\mathrm{post}}^{\mathrm{CNM}}$ or $\Gamma_{h,\mathrm{post}}^{\mathrm{EEM}}$, as the error covariance. Our metric for error is again the relative error
\begin{align}
E_{\mu_{\mathrm{a}}} = 100\%\cdot \frac{\|\mu_{\mathrm{a},\mathrm{true}} - \mu_{\mathrm{a},\mathrm{MAP}}\|}{\|\mu_{\mathrm{a},\mathrm{true}}\|}
\end{align}
where $\mu_{\mathrm{a},\mathrm{MAP}}$ denotes the MAP estimate of $\mu_{\mathrm{a}}$ and $h_{\mathrm{a},\mathrm{true}}$ denotes the true absorbed energy density shown in Figure \ref{FigureTruemu_a}.\par
\begin{figure}[H]
	\centering
	\begin{tabular}{c c c}
		\centering
		\includegraphics{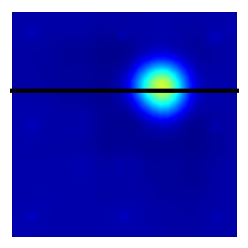} & 
		\includegraphics{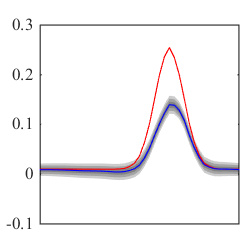}\\
		\includegraphics{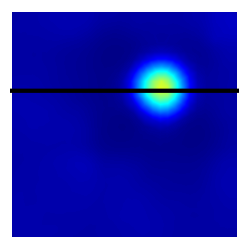} &
		\includegraphics{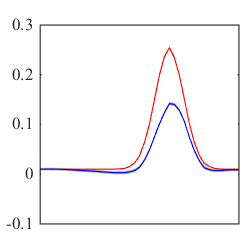}\\
		\includegraphics{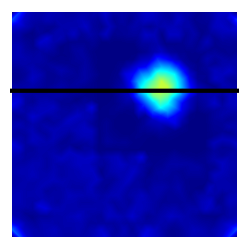}&
		\includegraphics{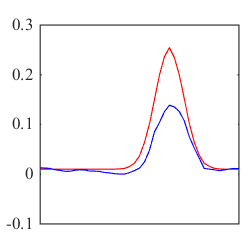}
	\end{tabular} 
	\caption{Conventional noise model posterior model of $\mu_{\mathrm{a}}$ using the conventional noise model MAP estimates $h_{\mathrm{MAP}}^{\mathrm{CNM}}$ of $h$ as data. Left column: MAP estimates $\mu_{\mathrm{a},\mathrm{MAP}}^{h\mathrm{CNM-CNM}}$. Right column: uncertainty estimates $\Gamma_{\mu_{\mathrm{a}},\mathrm{post}}^{h\mathrm{CNM-CNM}}$ with 1, 2 and 3 standard deviations displayed by the shading. The red line represents the true values and the blue line represents the MAP estimate values. Rows: noise levels of $5\%,1\%,0.1\%$ respectively.}
	\label{FigureOptConvNoiseReconsConv}
\end{figure}
\begin{figure}[H]
	\centering
	\begin{tabular}{c c c}
		\centering
		\includegraphics{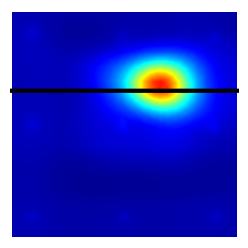} & 
		\includegraphics{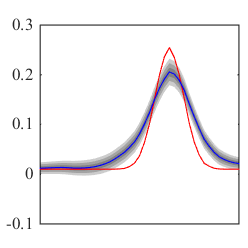}\\
		\includegraphics{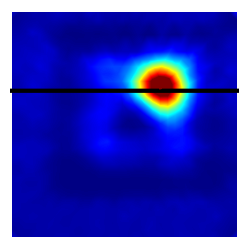} &
		\includegraphics{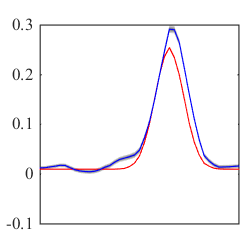}\\
		\includegraphics{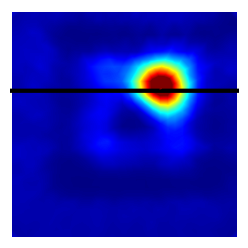}&
		\includegraphics{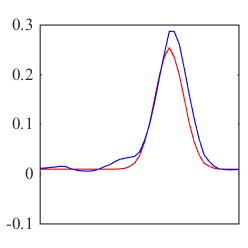}
	\end{tabular} 
	\caption{Conventional noise model posterior model of $\mu_{\mathrm{a}}$ using the enhanced error model MAP estimates $h_{\mathrm{MAP}}^{\mathrm{EEM}}$ of $h$ as data. Left column: MAP estimates $\mu_{\mathrm{a},\mathrm{MAP}}^{h\mathrm{EEM-CNM}}$. Right column: uncertainty estimates $\Gamma_{\mu_{\mathrm{a}},\mathrm{post}}^{h\mathrm{EEM-CNM}}$ with 1, 2 and 3 standard deviations displayed by the shading. The red line represents the true values and the blue line represents the MAP estimate values. Rows: noise levels of $5\%,1\%,0.1\%$ respectively.}
	\label{FigureOptEnhNoiseReconsConv}
\end{figure}

We begin by discussing the posterior models resulting from using the conventional noise model. In Table \ref{TableOpticalInverseErrorsConv}, we display the relative errors for conventional noise model MAP estimates of $\mu_{\mathrm{a}}$ obtained from using the conventional noise model $h_{\mathrm{MAP}}^{\mathrm{CNM}}$ and enhanced error model $h_{\mathrm{MAP}}^{\mathrm{EEM}}$ MAP estimates of $h$ as data. We denote these MAP estimates as $\mu_{\mathrm{a},\mathrm{MAP}}^{h\mathrm{CNM-CNM}}$ and $\mu_{\mathrm{a},\mathrm{MAP}}^{h\mathrm{EEM-CNM}}$ respectively. Further, we denote the corresponding uncertainty estimates as $\Gamma_{\mu_{\mathrm{a}},\mathrm{post}}^{h\mathrm{CNM-CNM}}$ and $\Gamma_{\mu_{\mathrm{a}},\mathrm{post}}^{h\mathrm{EEM-CNM}}$ respectively. In Figures \ref{FigureOptConvNoiseReconsConv} and \ref{FigureOptEnhNoiseReconsConv}, we display the MAP estimates and uncertainty estimates for these two cases.\par
In Figure \ref{FigureOptConvNoiseReconsConv}, it is clear that the true values are underestimated by the MAP estimates $\mu_{\mathrm{a},\mathrm{MAP}}^{h\mathrm{CNM-CNM}}$ and there is also a dip in the background values of $\mu_{\mathrm{a}}$ resembling the interior of the elastic layer. Furthermore, the uncertainty estimates $\Gamma_{\mu_{\mathrm{a}},\mathrm{post}}^{h\mathrm{CNM-CNM}}$ are extremely small and barely visible for the noise levels of $1\%$ and $0.1\%$. This is to be expected since the optical inverse problem possesses full domain data; every node on the discretized domain acts as a sensor. Furthermore, these error estimates show that our estimates are not feasible. This is also the case for Figure \ref{FigureOptEnhNoiseReconsConv}, where we show the uncertainty estimates $\Gamma_{\mu_{\mathrm{a}},\mathrm{post}}^{h\mathrm{EEM-CNM}}$ obtained from using the enhanced error model MAP estimates $h_{\mathrm{MAP}}^{\mathrm{EEM}}$ of $h$ as data. However, here we notice that the true values of the absorption coefficient are not as underestimated for the noise level of $5\%$ and even overestimated for the noise levels of $1\%$ and $0.1\%$.\par
\begin{figure}[H]
	\centering
	\begin{tabular}{c c c}
		\centering
		\includegraphics{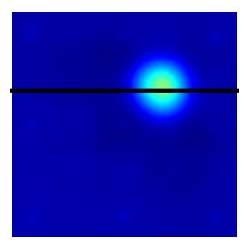} & 
		\includegraphics{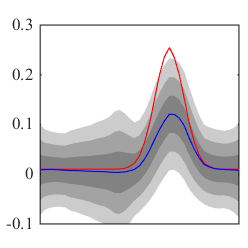}\\
		\includegraphics{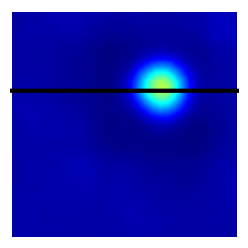} &
		\includegraphics{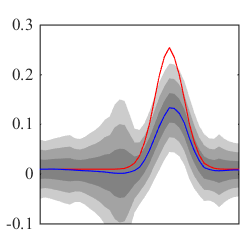}\\
		\includegraphics{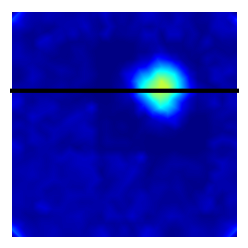}&
		\includegraphics{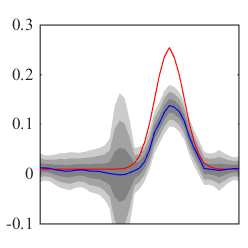}
	\end{tabular} 
	\caption{Coupled error model posterior model of $\mu_{\mathrm{a}}$ using the conventional noise model MAP estimates $h_{\mathrm{MAP}}^{\mathrm{CNM}}$ of $h$ as data and the corresponding posterior covariance $\Gamma_{h,\mathrm{post}}^{\mathrm{CNM}}$ as the error model covariance. Left column: MAP estimates $\mu_{\mathrm{a},\mathrm{MAP}}^{h\mathrm{CNM-CEM}}$. Right column: uncertainty estimates $\Gamma_{\mu_{\mathrm{a}},\mathrm{post}}^{h\mathrm{CNM-CEM}}$ with 1, 2 and 3 standard deviations displayed by the shading. The red line represents the true values and the blue line represents the MAP estimate values. Rows: noise levels of $5\%,1\%,0.1\%$ respectively.}
	\label{FigureOptConvNoiseReconsCoup}
\end{figure}
\begin{figure}[H]
	\centering
	\begin{tabular}{c c c}
		\centering
		\includegraphics{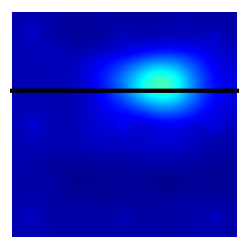} & 
		\includegraphics{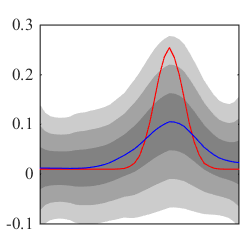}\\
		\includegraphics{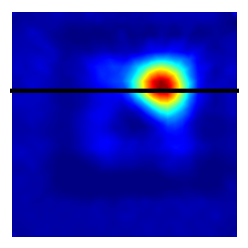} &
		\includegraphics{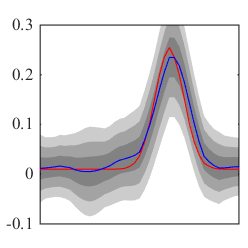}\\
		\includegraphics{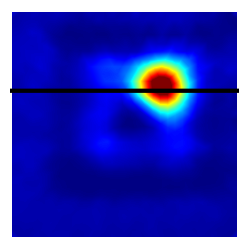}&
		\includegraphics{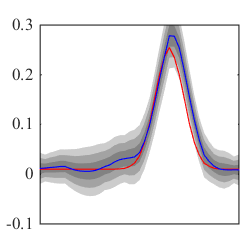}
	\end{tabular} 
	\caption{Coupled error model posterior model of $\mu_{\mathrm{a}}$ using the enhanced error model MAP estimates $h_{\mathrm{MAP}}^{\mathrm{EEM}}$ of $h$ as data and the corresponding posterior covariance $\Gamma_{h,\mathrm{post}}^{\mathrm{EEM}}$ as the error model covariance. Left column: MAP estimates $\mu_{\mathrm{a},\mathrm{MAP}}^{h\mathrm{EEM-CEM}}$. Right column: uncertainty estimates $\Gamma_{\mu_{\mathrm{a}},\mathrm{post}}^{h\mathrm{EEM-CEM}}$ with 1, 2 and 3 standard deviations displayed by the shading. The red line represents the true values and the blue line represents the MAP estimate values. Rows: noise levels of $5\%,1\%,0.1\%$ respectively.}
	\label{FigureOptEnhNoiseReconsCoup}
\end{figure}

\begin{table}
	\renewcommand{\arraystretch}{1.3}
	\caption{Relative errors for conventional noise model MAP estimates from the optical inverse problem.}
	\label{TableOpticalInverseErrorsConv}
	\centering
	\begin{tabular}{c | c | c | c}
		\centering
		& 5\% & 1\% & 0.1\% \\
		\hline
		Conventional: $\mu_{\mathrm{a},\mathrm{MAP}}^{h\mathrm{CNM-CNM}}$ & 40.621 & 42.6154 & 46.507\\
		Enhanced: $\mu_{\mathrm{a},\mathrm{MAP}}^{h\mathrm{EEM-CNM}}$ & 42.7119 & 43.0587 & 43.5029
	\end{tabular}
\end{table}
We now discuss the posterior models resulting from using the coupled error model. In Table \ref{TableOpticalInverseErrorsCoup}, we display the relative errors for coupled error model MAP estimates of $\mu_{\mathrm{a}}$ obtained from using the conventional noise model $h_{\mathrm{MAP}}^{\mathrm{CNM}}$ and enhanced error model $h_{\mathrm{MAP}}^{\mathrm{EEM}}$ MAP estimates of $h$ as data and their respective posterior covariances $\Gamma_{h,\mathrm{post}}^{\mathrm{CNM}}$ or $\Gamma_{h,\mathrm{post}}^{\mathrm{EEM}}$  as the error model covariance as discussed in Section \ref{SecOpticalInverseProblem}. We denote these MAP estimates as $\mu_{\mathrm{a},\mathrm{MAP}}^{h\mathrm{CNM-CEM}}$ and $\mu_{\mathrm{a},\mathrm{MAP}}^{h\mathrm{EEM-CEM}}$ respectively. Further, we denote the corresponding uncertainty estimates as $\Gamma_{\mu_{\mathrm{a}},\mathrm{post}}^{h\mathrm{CNM-CEM}}$ and $\Gamma_{\mu_{\mathrm{a}},\mathrm{post}}^{h\mathrm{EEM-CEM}}$ respectively. In Figures \ref{FigureOptConvNoiseReconsCoup} and \ref{FigureOptEnhNoiseReconsCoup}, we display the MAP estimates and uncertainty estimates for these two cases.\par
In Figure \ref{FigureOptConvNoiseReconsCoup}, it is clear that the true values are underestimated by $\mu_{\mathrm{a},\mathrm{MAP}}^{h\mathrm{CNM-CEM}}$ and there is also a dip in the background values resembling the interior of the elastic layer. Furthermore, the coupled error model uncertainty estimates $\Gamma_{\mu_{\mathrm{a}},\mathrm{post}}^{h\mathrm{CNM-CEM}}$ inherits the shape of the uncertainty estimates of the posterior model covariance $\Gamma_{h,\mathrm{post}}^{\mathrm{CNM}}$ of $h$ obtained using the conventional noise model in the acoustic inverse problem. These show that our estimates are infeasible. In contrast, we can see in Figure \ref{FigureOptEnhNoiseReconsCoup} that using the posterior model covariance $\Gamma_{h,\mathrm{post}}^{\mathrm{EEM}}$ of $h$ obtained from using the enhanced error model yields feasible estimates of the absorption coefficient without a spike in the uncertainty estimates. Overall, we can see that using the posterior model for $h$ obtained from using the enhanced error model yields a more accurate posterior model for $\mu_{\mathrm{a}}$. This observation is also quantitatively supported by the relative errors displayed in Table \ref{TableOpticalInverseErrorsCoup}.
\begin{table}
	\renewcommand{\arraystretch}{1.3}
	\caption{Relative errors for coupled error model MAP estimates from the optical inverse problem.}
	\label{TableOpticalInverseErrorsCoup}
	\centering
	\begin{tabular}{c | c | c | c}
	\centering
	& 5\% & 1\% & 0.1\% \\
	\hline
	Conventional: $\mu_{\mathrm{a},\mathrm{MAP}}^{h\mathrm{CNM-CEM}}$ & 45.9753 & 44.3749 & 45.767\\
	Enhanced: $\mu_{\mathrm{a},\mathrm{MAP}}^{h\mathrm{EEM-CEM}}$ & 51.0612 & 35.5756 & 42.1139
	\end{tabular}
\end{table}
\section{Conclusions}
For the acoustic inverse problem, the results show that utilizing the enhanced error model to account for the modeling errors arising from neglecting the elastic error yields a more accurate posterior density. For the optical inverse problem, the results show that utilizing the conventional noise model yields infeasible uncertainty estimates even when the improved enhanced error model MAP estimates of $h$ are used as data. In particular, this leads to unrealistically certain belief in quality of our absorption estimates. With the coupled error model, only when the enhanced error model MAP estimates of $h$ is used as data are feasible estimates obtained. In conclusion, both the enhanced error model for the acoustic inverse problem and the coupled error model for the optical inverse problem are required to obtain feasible estimates of the absorption coefficient distribution.\par
Overall, the numerical results in this work demonstrate that the accuracy of the posterior model can be improved by using Bayesian approximation errors to account for the modelling errors arising from inaccurate modelling of the elastic layer in the inversion process. Furthermore, the inclusion of this posterior model in the error model of the inversion process for the optical inverse problem can improve the feasibility of the estimates of the absorption coefficient distribution.

\section*{Acknowledgment}
The Dodd-Walls Centre PhD scholarship, Academy of Finland (projects 286247 and 312342) and Jane and Aatos Erkko Foundation.

\nocite{*}
\bibliography{references}{}
\bibliographystyle{IEEEtran}

\end{document}